\documentclass[aps, prd, reprint, superscriptaddress, longbibliography, nofootinbib, floatfix]{revtex4-2}

\usepackage{ragged2e}
\usepackage{slashed}
\usepackage{verbatim}
\usepackage[T1]{fontenc}
\usepackage[utf8]{inputenc}
\usepackage[colorlinks]{hyperref}
\usepackage{mathbbol}
\usepackage[dvipsnames]{xcolor}
\usepackage{graphicx}
\usepackage{orcidlink}
\usepackage{euscript}
\usepackage[normalem]{ulem}
\usepackage{svg}
\usepackage[english]{babel}
\usepackage{lipsum}
\usepackage{physics}
\usepackage{dcolumn}
\usepackage{tensor}
\usepackage{comment}
\usepackage{graphicx,color,overpic,mathtools}
\usepackage{amsthm,amsmath,amssymb,hyperref,mathrsfs}
\usepackage{amssymb}
\usepackage{braket,bm,bbm,setspace}
\usepackage{cancel}
\usepackage{float}
\usepackage{realhats}
\usepackage{galdrastafir}

\hypersetup{
    colorlinks=false,
    pdfborder={0 0 0},
}
\usepackage{xargs}

\colorlet{RED}{red}

\usepackage{staves}

\newcommand{\marksize}{\fontsize{4}{5}\selectfont}

\makeatletter
\def\@fnsymbol#1{%
  \ifcase#1\or
    \hulinhjalmur[size=1.1em]\or
    \marksize\staveI\or
    \marksize\staveXXIV\or
    \marksize\staveV\or
       
    \marksize\staveII\or
    \marksize\staveVI\or
    \marksize\staveVII\or
    \marksize\staveVIII\or
    
    \marksize\staveIII\or
    
    \marksize\staveIV\or
    \marksize\staveIX\else
    
    \@ctrerr
  \fi
}
\makeatother

\usepackage{calc}
\newlength{\orcidwd}
\newlength{\hatwd}
\newcommand{\orcidhat}[2]{%
  \settowidth{\orcidwd}{\orcidlink{#2}}%
  \settowidth{\hatwd}{\hat[#1]{\orcidlink{#2}}}%
  \makebox[\orcidwd]{\hat[#1]{\orcidlink{#2}}}%
  \kern\dimexpr(\hatwd-\orcidwd)/2\relax
}

\definecolor{mypurple}{RGB}{130, 0, 130} 
\definecolor{oxblood}{rgb}{0.5333, 0.0314, 0.0314}
\definecolor{orange}{rgb}{1,0.5,0}

 \hypersetup{
    colorlinks=true,
    linkcolor=oxblood, 
     citecolor=oxblood, 
    urlcolor=oxblood   
 }
 
\begin{document}

\title{Strong-Field Gravitational-Wave Lensing from Ray-Traced Arrival-Time Surfaces}

\author{Luka Vujeva\,\orcidlink{0000-0001-7697-8361}}
\email{luka.vujeva@nbi.ku.dk}
\affiliation{Center of Gravity, Niels Bohr Institute, Blegdamsvej 17, 2100 Copenhagen, Denmark}

\author{Jo\~ao S. Santos\,\orcidlink{0000-0003-1921-8758}}
\email{joao.sieiro.santos@tecnico.ulisboa.pt}
\affiliation{Center of Gravity, Niels Bohr Institute, Blegdamsvej 17, 2100 Copenhagen, Denmark}
\affiliation{CENTRA, Departamento de Física, Instituto Superior Técnico -- IST,
Universidade de Lisboa -- UL, Avenida Rovisco Pais 1, 1049-001 Lisboa, Portugal}

\author{Brian C. Seymour\,\orcidlink{0000-0002-7865-1052}}
\email{brian.seymour@nbi.ku.dk}
\affiliation{Center of Gravity, Niels Bohr Institute, Blegdamsvej 17, 2100 Copenhagen, Denmark}




\begin{abstract}

 The diffraction integral formalism has been greatly successful in describing gravitational wave lensing effects in scenarios where the wavelength of the signal is comparable to the characteristic length scale of the lens, known as the wave optics regime. 
 However, the assumptions required to construct its integrand break down as the source approaches the strong gravity region of a black hole lens. 
 In particular, we are interested in the lensed signal from a binary (the source) near a massive black hole (the lens).
 We present a new formalism for obtaining low frequency corrections to sources being lensed in the strong field regime of gravity by leveraging the fact that null geodesics obey the Hamilton--Jacobi equations. 
 In doing so, we are able to construct a time delay surface without relying on the thin lens approximation, and solve the diffraction integral for the resulting surface directly. 
For a stationary source at $10M$, deep in the strong field regime, when comparing the results with a relativistic wave scattering model, we find $6.5\%$ and $0.3\%$ agreement in the peak amplification and first diffractive fringe, respectively. 
 Furthermore, we find a $< 1\%$ agreement with the amplitude of the point mass lens with a source at $10^4 M$ in the weak field, but noticeable dephasing between the peaks of the amplification function at higher frequencies.
 This work allows for clean interpolation between otherwise separate regimes, valid for sources from $\sim 6M \to \infty$. 
 It provides a transparent, physically motivated framework for computing beyond-geometric optics lensing effects in the strong field regime.
\end{abstract}

\maketitle

%
\section{Introduction}

Gravitational wave (GW) lensing offers a unique probe into the dark Universe, particularly in the regime in which the wavelength of the GW approaches the characteristic size of the lens (known as the wave optics (WO) regime). There has been a concerted effort in extending the language of standard electromagnetic lensing to the WO regime using traditional Kirchhoff diffraction integral approaches~\cite{Schneider:1992bmb, Deguchi:1986zz, Nakamura:1999uwi, Takahashi:2003ix, Tambalo:2022plm, Villarrubia-Rojo:2024xcj}. However, the standard approximations taken to obtain the frequency dependent amplification function, such as the geometric optics (GO) and thin lens limits, need not apply to a GW lensing system. A prime example is a binary system evolving near a massive black hole (BH), commonly referred to as binary extreme mass ratio inspirals~\cite{Bartos:2016dgn,Stone:2016wzz,Dittmann:2023sha, Peng:2024wqf,Vaccaro:2025dzl,Tagawa:2019osr,Samsing:2020tda, Hoang:2017fvh, Chen:2018axp}. These triple systems carry signatures of their dynamical environment~\cite{Barausse:2014tra, Meiron:2016ipr, Inayoshi:2017hgw,Yu:2020dlm,Laeuger:2023qyz, Yin:2024nyz, Zwick:2025wkt, Zwick:2025ine,Zwick:2026thw, Santos:2025ass}, one of them being gravitational lensing. Using standard GO and thin lens approximations, these systems have been amply explored in the literature~\cite{DOrazio:2017ssb, DOrazio:2019fbq,Yu:2021dqx, Leong:2024nnx, Ubach:2025dob, Ubach:2025oxr, Martens:2026mku}. However, it is known that this setup may break these approximations~\cite{Suyama:2005mx, Guo:2024hmp, Pijnenburg:2024btj, Chan:2025wgz}. Consequently, there has been a great deal of work to include finite wavelength effects~\cite{Harte:2018wni, Cusin:2019rmt, Ezquiaga:2025gkd, Oancea:2022szu, Kubota:2024zkv, Pijnenburg:2024btj, Guo:2024hmp, Nazare:2026glo, Cardoso:2026ugm}, as well as going beyond the thin lens approximation~\cite{Suyama:2005mx, Feldbrugge:2020ycp, Ramesh:2021nnl, Mizuno:2022xxp, Carrillo-Gonzalez:2026oov,Huang:2026hbb}. A complementary approach that is manifestly WO and thick lens is to work with partial waves adapted to the massive BH spacetime~\cite{Nambu:2012wa, Folacci:2019vtt, Chan:2025wgz, Li:2025gou,  Saketh:2025cwf, Motohashi:2021zyv}.

In this paper, we present yet another approach  to building towards an amplification function in the strong field regime, in the high (but finite) frequency limit. We use numerical ray tracing techniques to construct the analog of the weak field arrival time surface, allowing us to obtain lower frequency corrections to the standard stationary phase approximation.
Our approach is intuitive and transparent: it assumes radiation travels along null rays but keeps the full information of their relative phases. We compare it with WO and GO results for sources near the optical axis, as a step toward connecting the strong and weak field regimes.

We start with a review of the standard weak field diffraction integral formalism in Sec.~\ref{sec:diff_int}, including a discussion  of our strong field setup's departure from its realm of validity. 
In Sec.~\ref{sec:td_surface} we detail the construction of the strong field time delay surface from Hamilton--Jacobi theory, and in Sec.~\ref{sec:amp_f} we use it to obtain the  amplification function.
 Additionally, we recover similar notions to the arrival times, magnifications, and image parities seen in the standard diffraction integral literature, useful in proving the consistency of this setup with existing works. In Sec.~\ref{sec:results}, we show comparisons to full wave scattering simulations in Schwarzschild to validate the setup against methods of describing these systems fully self consistently in the strong field (cf.~Appendix~\ref{app:scalar_WO}). We summarize and outline future research directions in Sec.~\ref{sec:conclusions}. From Sec.~\ref{sec:td_surface} onward we use units where $G=c=1$. 
%
%
\section{The Diffraction Integral}
\label{sec:diff_int}

In this section, we provide an overview of the weak field diffraction integral formalism, and highlight the key areas where it breaks down in the context of approaching the vicinity of a strong field lens.

\subsection{The Weak Field Diffraction Integral}
\label{subsec:weakfield_lensing}
Lensing of GWs is typically split into two distinct regimes: GO
when the wavelength of the signal is much smaller than the Schwarzschild radius of the
lens (i.e. $\lambda_{\rm GW} \ll R_{\rm S}$), and WO when the wavelength
is comparable ($\lambda_{\rm GW} \sim R_{\rm S}$). In the GO regime, the lensing
effects are achromatic. 
In the WO regime, diffraction makes the lensing frequency dependent, which is described by the amplification function
\begin{equation}
    F(f) = \frac{ \tilde h_L(f)}{\tilde h_0(f)} ,\label{eq:F_f}
\end{equation}
where $\tilde h_L(f)$ and $\tilde h_0(f)$ are the lensed and unlensed GW strains in the frequency domain. 
For a fixed weak-field lens model (where the lens is well described by a Newtonian potential) and distance scaling, such as the Einstein angle, the source position enters through its two-dimensional angular position on the sky.
This relies on the thin-lens approximation, in which the deflection is taken
to occur instantaneously in a single plane transverse to the line of sight, so that
the geometry of the problem reduces to a map between two flat planes and the radial
extent of the deflecting region is discarded entirely. The distance along the line of
sight at which the source sits therefore never appears on its own; it enters only
through the angular diameter distances that set the overall scale. As we show in
Sec.~\ref{sec:td_surface}, it is precisely this reduction that fails for a source in
the immediate vicinity of a black hole, where the deflection accumulates over a radial
range comparable to the source--lens separation itself, and where the radial position
$r_{\rm S}$ is consequently a physical parameter of the problem rather than a scale
that can be absorbed.

In the WO regime, the strain of the lensed GW is determined by the Kirchhoff diffraction
integral~\cite{Schneider:1992bmb, Takahashi:2003ix, Tambalo:2022wlm}, given by
\begin{equation}
    F(w;\vec{y}) =  \frac{w}{2\pi i} \int d^2 \vec{x}\;
    e^{iw T_d(\vec{x}, \vec{y})}, \label{eq:kirchhoff}
\end{equation}
where $T_d(\vec{x}, \vec{y})$ is the Fermat potential, and $w$ is the
dimensionless frequency defined as $w \equiv 8 \pi G M_{Lz} f$, where $G$ is
Newton's constant. The redshifted lens mass $M_{Lz}$ is defined as
\begin{equation}
    M_{Lz} \equiv (1+z_L)\frac{D_S}{D_L D_{LS}}  \frac{\theta_E^2}{4G} \, ,
\end{equation}
where $z_L$ is the lens redshift, $D_S$, $D_L$, and $D_{LS}$ are observer--source,
observer--lens and lens--source angular diameter distances, respectively, and
\begin{equation}
    \theta_E \equiv \sqrt{4 G M (D_L D_{LS}/D_S)}
    \label{eq:einstein_radius}
\end{equation}
is the Einstein angle. In Eq.~\eqref{eq:kirchhoff} the image-plane and source-plane positions have been rescaled by the angle $\theta_E$, as
\begin{equation}
    \vec{x} = \frac{\vec\theta}{\theta_E}\,, \qquad
    \vec{y} = \frac{\vec\beta}{\theta_E}\,,
    \label{eq:dimensionless_coords}
\end{equation}
where $\vec\theta$ and $\vec\beta$ are the angular image and source positions, respectively. In
these units the Fermat potential, also called the time delay surface \cite{Blandford:1986zz, Schneider:1992bmb}, can be written as
\begin{equation}
    T_d(\vec{x},\vec{y}) \equiv  \bigg[  \frac{(\vec{x} - \vec{y})^2}{2}
    - \psi(\vec{x}) \bigg ], \label{eq:td_weak_field}
\end{equation}
where $\psi(\vec{x})$ is the Newtonian lens potential. For a given lensing configuration in GO,
the locations of images are determined by the lens equation,
\begin{equation}
    \vec{y} = \vec{x} - \vec\alpha(\vec{x}), \label{eq:lensequation}
\end{equation}
where $\vec{y}$ is the source location, $\vec{x}$ is the image location, and
$\vec\alpha(\vec{x})$ is the deflection angle, determined by the lensing potential as
$\vec\alpha(\vec{x}) = \nabla \psi(\vec{x})$. An important note for the later sections
of this work is that this construction rests on an expansion in small deflection
angles, in addition to the thin-lens reduction above. Neither is guaranteed to hold when the source is placed in the immediate vicinity of a black hole lens, and our beyond geometric optics approach aims to address this issue. 
In this framework, the images occur at stationary points of the time delay surface, i.e. where $\nabla_{\vec{x}} T_d = 0$. 

\begin{figure*}
    \centering
    \includegraphics[width=\linewidth]{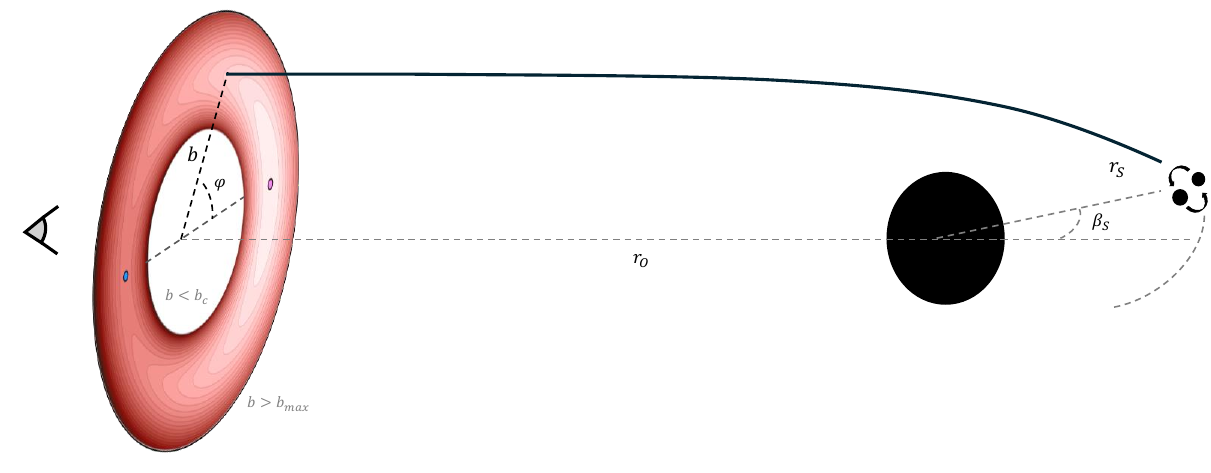}
    \caption{The geometric setup considered in this work. The observer is placed at a distance $r_{\rm O}$ from the center of the lens. The source is placed at a radius $r_{\rm S}$ from the lens, at an angular offset of $\beta_{\rm S}$ from the antipodal point. Geodesics are traced from observer screen positions $\vec\theta = (b, \varphi)$ back to the source radius. 
    }
    \label{fig:setup}
\end{figure*}

\subsection{The point mass lens}
\label{subsec:PL}

The weak-field counterpart of the black hole lens is the point mass lens (PL),
obtained by retaining only the leading order term in the expansion of the deflection
in $M/b$, $\alpha(b) = 4M/b$, and collapsing the deflection onto a single plane
transverse to the line of sight 
It is the model that is typically adopted to
describe WO effects in the GW lensing literature, and in
particular, it is the model applied to lensing by the supermassive black holes at the
centers of AGNs~\cite{Leong:2024nnx, Ubach:2025dob, Ubach:2025oxr, Martens:2026mku}.
That configuration is precisely the one a binary extreme mass ratio inspiral realizes. The important
difference is that the emitting binary sits at $\mathcal{O}(10-100)M $ rather than rather than at the typical cosmological distances considered in thin-lens setups. We therefore
adopt the PL as one of the two benchmarks against which the formalism developed here
is tested, the other being the fully relativistic WO scalar scattering (see Appendix~\ref{app:scalar_WO}). The two bracket the construction from either side, in
that the PL is the weak-field limit that our surface must reproduce when the source is
moved far from the lens, while the scattering calculation is the exact result that it
must approach in the strong field, and establishing where the PL departs from the
latter is a direct statement about the reliability of the forecasts already in the
literature.

Following the derivations of Refs.~\cite{Schneider:1992bmb, Takahashi:2003ix}, we work with the dimensionless coordinates of Eq.~\eqref{eq:dimensionless_coords}, and note that the spherical symmetry of the lens model allows us to work with a single radial coordinate on either the source or image plane. Moreover, all other quantities defined in the remainder of this section are also dimensionless: angles are measured in units of the Einstein radius, and arrival times in units of $4GM_{Lz}/c^3$, so that a physical time delay is recovered as $\Delta t = 4GM_{Lz}\Delta T_d/c^3$. 
In these units the PL potential is simply
\begin{equation}
    \psi(x) = \ln|x| \,, \label{eq:lens_potential_PL}
\end{equation}
so that the time delay surface of Eq.~\eqref{eq:td_weak_field} takes the familiar
two-term form of a quadratic geometric term minus the deflection potential,
\begin{equation}
    T_d(x,y)
    = \frac{1}{2}\left(x-y\right)^2
    - \ln|x|
    - \phi_m(y)\,,
    \label{eq:td_PL}
\end{equation}
where
\begin{equation}
        \phi_m(y) = \frac{\left(x_m - y\right)^2}{2} - \ln|x_m|
\end{equation}
is the reference phase evaluated at the minimum of the surface, $x_m$. 

A convenient property of the PL is that the lens equation, the image positions,
their magnifications $\mu$, and their arrival times all admit closed forms in terms of the source
position alone, 
\begin{equation}
    y = x - \frac{1}{x}\,,
    \qquad
    x_\pm = \frac{1}{2}\left( y \pm \sqrt{y^2+4} \right)\,,
\end{equation}
\begin{equation}
    \mu_\pm = \pm\frac{1}{2} + \frac{y^2+2}{2y\sqrt{y^2+4}}\,,
\end{equation}
\begin{equation}
    \Delta T_d(y) = \frac{y\sqrt{y^2+4}}{2}
    + \ln\frac{\sqrt{y^2+4}+y}{\sqrt{y^2+4}-y}\,,
\end{equation}
with $x_m = x_+$ the minimum of Eq.~\eqref{eq:td_PL}, and where the $\mu_\pm$ are
magnifications in absolute value, so that the parity of the saddle image must be
tracked separately. These quantities are commonly used to search for overlapping
lensed images in current lensed GW search pipelines~\cite{Chan:2024qmb, Chan:2025kyu,
Chan:2025pdf, LIGOScientific:2021izm, LIGOScientific:2023bwz, LIGOScientific:2025cwb}.

The
GO description above breaks down when $\lambda_{\rm GW}\sim R_{\rm
S}$, and the full diffraction integral must be evaluated instead. Thankfully, the PL
is one of very few known Newtonian lens models with a closed form amplification
function~\cite{Takahashi:2003ix}, which can be written in terms of gamma and
hypergeometric functions as
\begin{equation}
\begin{aligned}
F(w;y)
&= \exp\!\left[
\frac{\pi w}{4}
+\frac{iw}{2}
\left(\ln\frac{w}{2}-2\phi_m(y)\right)
\right] \\
&\quad \times
\Gamma\!\left(1-\frac{iw}{2}\right)
{}_1F_1\!\left(
\frac{iw}{2};\,1;\,\frac{iw y^2}{2}
\right),
\end{aligned}
\end{equation}
or equivalently in terms of Laguerre polynomials~\cite{Ezquiaga:2020spg}. It is this
expression that we compare against throughout.

Three assumptions underlying this construction fail once the source is placed in the
immediate vicinity of the lens. The first two are the thin lens and small deflection
angle approximations discussed in Sec.~\ref{subsec:weakfield_lensing}, which enter
here through Eq.~\eqref{eq:lens_potential_PL} and, in our presentation, through
Eqs.~\eqref{eq:td_weak_field} and \eqref{eq:lensequation}.
Their most visible casualty is the winding structure of the strong field, since a
surface built on a single-plane deflection admits no images formed by geodesics that
cross the plane perpendicular to the optical axis and passing through the origin more than once~\cite{Bozza:2002zj, Bozza:2001xd,
Bozza:2010xqn, Vazquez:2003zm, Gralla:2019drh, Gralla:2019ceu}. 
The third is that sources are assumed to be stationary, point-like and isotropic emitters, which is not representative of a compact binary emitter with its distinct quadrupolar emission pattern~\cite{Yu:2021dqx}. A detailed exploration of these assumptions can be found in Refs.~\cite{Suyama:2005mx,
Bruyere:2026gnt, Bruyere:2026xzb}.

Despite these differences, the amplification function of Eq.~\eqref{eq:F_f} is defined as a ratio of strains, and therefore requires no thin-lens input of its own; what the weak-field machinery supplies is not the definition, but a means of computing it. In the strong field, Fermat's principle motivates constructing an arrival-time surface, whose stationary points should give the connecting rays and their travel times. Away from those
points, the surface is instead a phase entering the diffraction integral, and not the travel time of any ray connecting the source and observer.

Replacing the surface alone does not, however, fix the weight of each contribution, which also depends on how the rays are focused. A noteworthy caveat is that the resulting construction is eikonal throughout, so the only frequency dependence it can generate is that arising from the interference of rays.

Crucially, this extension provides a notion of the arrival times of discrete images in
the high frequency limit, and provides deeper insight into the interference structure
found in the full WO solutions (cf.~Fig.~1 in Ref.~\cite{Santos:2025ass}).
While we consider a static, approximately monochromatic source in this configuration,
insights from the high-frequency limit that provide a well defined image arrival time,
phase, and image multiplicity can help understand some of the frequency dependent
oscillations found in lensed binary extreme mass ratio waveforms~\cite{Santos:2025ass, Santos:2026lzq}.

\begin{figure*}
    \centering
        \includegraphics[width=\linewidth]{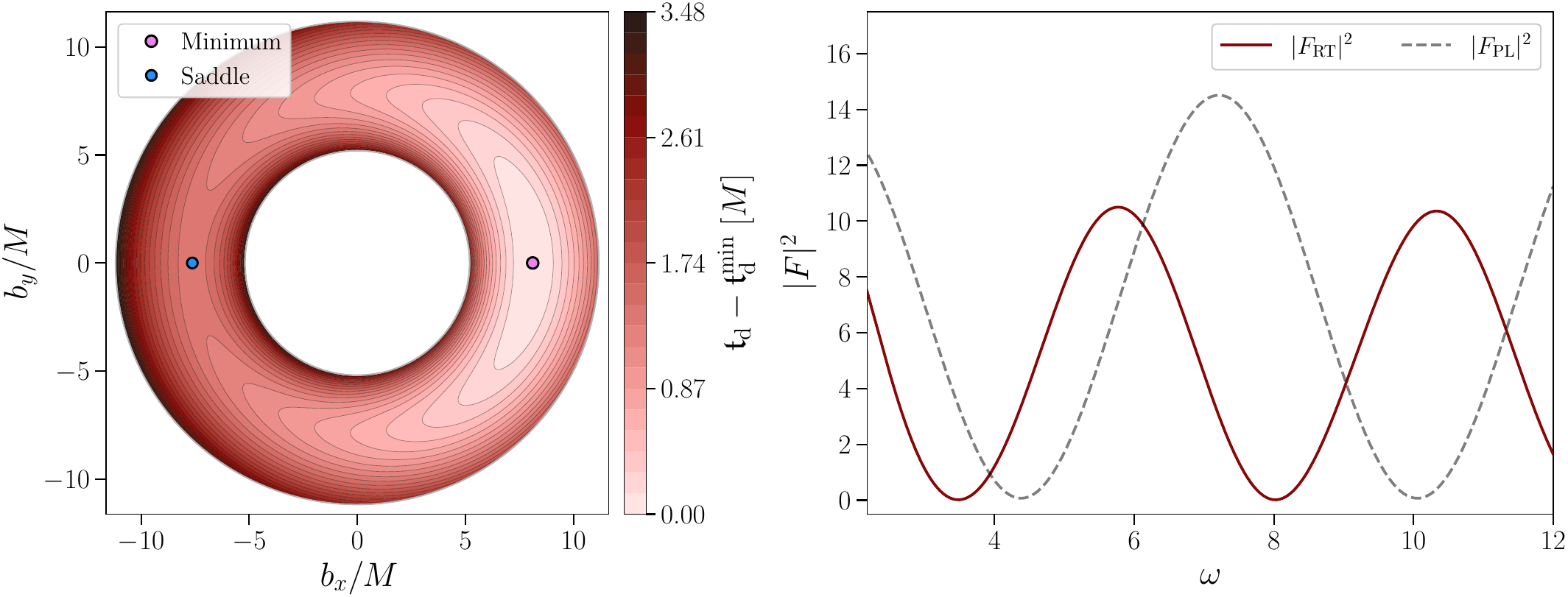}
    \caption{(Left) Arrival-time surface for the direct images, $\mathfrak{t}_{\mathrm{d},0} (b, \varphi, \beta_{\rm S})$, of a Schwarzschild lens with source radius $r_{\rm S}=10\,M$ and azimuthal location $\beta_{\rm S} = \pi/36$. We use Cartesian screen coordinates $b_x=b\cos\varphi$, $b_y=b\sin\varphi$ with $b=\sqrt{b_x^2+b_y^2}$. Contours of equal time delay are over-plotted, highlighting the recovery of the structure of the stationary points matching those of the weak field lensing literature. The images form at a minimum and a saddle point of the surface, corresponding to a type-I and a type-II image carrying the expected $\pi/2$ relative phase difference. (Right) Corresponding amplification function $|F_\mathrm{RT}|^2$, compared against a point mass lens $|F_\mathrm{PL}|^2$. 
    }
    \label{fig:td_schw}
\end{figure*}
\section{Construction of the Arrival-Time Surface}\label{sec:td_surface}
In this section, we outline the eikonal generating-function formalism used to assemble the arrival time surface. This is the centerpiece of our construction of a strong-field analogue of the amplification function $F(\omega)$ familiar from weak-field lensing (Eq.~\eqref{eq:F_f}). As outlined in Sec.~\ref{sec:diff_int}, the weak-field construction usually assumes small deflections, a thin Newtonian lens, and a fixed source--lens--observer geometry. Within this framework, Eq.~\eqref{eq:kirchhoff} computes the interference of distinct rays connecting source and observer. Our framework is based on identifying in the Fermat potential in Eq.~\eqref{eq:kirchhoff} $T_d$, the arrival time surface for these rays. So the statement 
\begin{eqnarray}
    F(\omega, \vec{\beta}) \sim \omega \int d^2 \vec{\theta} e^{i \omega \mathfrak{t}_{\rm{d}}(\vec{\theta}, \vec{\beta})} \, ,
\end{eqnarray}
is well defined in the strong field (the rigorous and useful definition is provided in Sec.~\ref{sec:amp_f}). We call the strong field quantity $\mathfrak{t}_{\rm{d}}(\vec{\theta}, \vec{\beta})$ to distinguish it from the weak field Fermat potential; in the weak field both results agree (see Sec.~\ref{sec:results}). The main difference is that rather than using a Newtonian lens potential ($\psi$ in the last section), we trace out individual null geodesics connecting source and observer to build $\mathfrak{t}_{\rm{d}}$. It represents the leading GO interference of null rays at some observer direction (which sets the optical axis) and coming from a specific source position.  

Our construction does not include any beyond-GO effects for the propagation of the field, namely the amplitude transport along the null geodesics (cf.~\cite{Bruyere:2026xzb,Cusin:2019rmt}). Indeed, all the finite frequency effects come from not using the stationary phase to solve these integrals. In other words, we are not making a consistent expansion in powers of $(M\omega)^{-1}$; instead, we keep some of the terms in this expansion (the interference integral), which suffice to capture the WO and finite lens effects. Physically, the waves are attached to rays, but we track their relative phase at the observer.

Our method begins with fixing the source position, as illustrated in Fig.~\ref{fig:setup}.
The observer is placed at a distance $r_{\rm O}$ from the center of the lens. The source is placed at a radius $r_{\rm S}$ from the lens, at a \emph{small} angular offset $\beta_{\rm S}\ll1$ from the antipodal point. Here $r_{\rm S}$ and $r_{\rm O}$ are Schwarzschild radial coordinates.  Because the Schwarzschild geometry considered throughout this work is spherically symmetric, a single angle rather than a vector source-plane position specifies the configuration completely. Geodesics are then traced perpendicularly from observer screen positions $\vec\theta = (b, \varphi)$ back to the source radius, where $b$ and $\varphi$ are the screen impact-parameter and azimuth coordinates relative to $\beta_{\rm S}$, respectively. The range of impact parameters considered for our time delay surface is set out in Sec.~\ref{sec:td_domain}.  

We stress that, for a given observer direction, not all rays (in fact only two per source position per winding number, see Sec.~\ref{sec:td_winding}) connect to the source point. All the others will connect to nearby points at the same radius; this smudging of the source is standard in ray-tracing setups. Our strong field $\mathfrak{t}_{\rm d}$ must encode these connecting rays through its stationary points, just as we saw in Sec.~\ref{sec:diff_int}. The eikonal generating function, which we define below, turns out to be very close to being the correct object. Hereafter, $\omega$ is the angular frequency defined using Schwarzschild coordinate time. We express frequencies in dimensionless form as $M\omega$. In the absence of cosmological redshift, this is related to the usual weak-field frequency by $w=4M\omega$.
\subsection{The eikonal generating function}
\label{sec:td_generating}
The approach used in this section follows the Hamilton--Jacobi formulation of geodesic motion, in which derivatives of the separated action with respect to the conserved energy and angular momentum determine the coordinate travel time and azimuthal sweep, respectively~\cite{Carter:1968rr,Chandrasekhar:1985kt}. Analytic treatments of Schwarzschild trajectories and their associated times include the classic studies of Darwin~\cite{1959RSPSA.249..180D,Darwin:1961-gravityfieldII}, with the orbital solutions expressed in terms of elliptic functions. Here, we use the radial action as a generating function to relate the travel time directly to the azimuthal sweep. This provides a convenient way to reconstruct the travel time from the angular integral while retaining the exact Schwarzschild geodesics.

Null geodesics in Schwarzschild spacetime are confined to move on a plane containing the origin and, for a given asymptotic direction (in our case the optical axis), they are parameterized by a single parameter, the impact parameter $b\equiv L/E$ (see Fig.~\ref{fig:setup}), for $L$ and $E$ the null particle's angular momentum and energy, respectively. For any such geodesic, the (elliptic) radial leg integrals
\begin{eqnarray}
    \Phi_{\rm leg}(R;b)
&=&
\int_{r_t}^{R}
\frac{b/r^2}{\sqrt{\mathcal{R}(r;b)}}\,dr,
\qquad\label{eq:elliptic_phi}
\\
T_{\rm leg}(R;b)
&=&
\int_{r_t}^{R}
\frac{1}{(1-2M/r)\sqrt{\mathcal{R}(r;b)}}\,dr, \label{eq:elliptic_t}
\end{eqnarray}
with the Schwarzschild radial potential
\begin{equation}
\mathcal{R}(r;b)
=
1-\frac{b^2}{r^2}\left(1-\frac{2M}{r}\right) ,
\label{eq:Rradial}
\end{equation}
yield the accumulated azimuthal phase and coordinate time from an inner turning point $r_t$ out to some radius $R$. In a typical approach to the problem of timing geodesics, one solves the time integral once from the observer radius, down to a turning point, and a second time back out to the source radius. The azimuthal sweep $\Phi(b)$ is expressible in an incomplete elliptic integral of the first kind~\cite{1959RSPSA.249..180D,Darwin:1961-gravityfieldII,Chandrasekhar:1985kt}, whereas closed forms for the travel time $T(b)$ also involve elliptic integrals of the third kind and are considerably more cumbersome~\cite{Hackmann:2018fmk,Gralla:2019ceu}. 

Instead of integrating Eq.~\eqref{eq:elliptic_t} directly, we leverage the structure inherited by the radial action $\mathcal I(b)$ for null geodesics in the Hamilton--Jacobi equations~\cite{Carter:1968rr,Chandrasekhar:1985kt,Katanaev:2023qiv}. The radial action is also known as the eikonal generating function, which exists because of the symmetries of Schwarzschild.
In particular, we have
\begin{eqnarray}
    \mathcal{I}(R;b) 
    = \int_{r_t} ^R p_r\,\mathrm{d}r =  \int_{r_t} ^R  \frac{\sqrt{\mathcal{R}(r;b)}}{(1-2M/r)}\,dr\,,
\end{eqnarray}
for $p_r$ the canonical radial momentum. For our purposes, we are interested in studying $\mathcal I (b)$ at fixed $R$. This object has several important properties. First, taking a derivative with respect to $b$ yields\footnote{There is a small caveat: the turning point radius $r_t$ depends on the impact parameter, which means there could be a boundary term. However, this is not the case as $p_r (r-r_t) \sim (r-r_t)^{1/2} $.}
\begin{eqnarray}
    \mathcal{I}^\prime(b) = -\Phi(b) \, ,
\end{eqnarray}
which implies that $\mathcal I(b)$ decreases with $b$ for positive $\Phi$. This is independent of the angular source displacement, but depends on both the source and observer radii through the radial integrals. On the other hand, it is easy to prove that 
\begin{eqnarray}
    \mathcal I (b) = T(b) - b \Phi (b) \, , \label{eq:generating}
\end{eqnarray}
which is an exact consequence of the Hamilton--Jacobi structure of the null geodesics of Schwarzschild. Combining both results yields the fundamental relation 
\begin{equation}
    \frac{\mathrm{d}T}{\mathrm{d}\Phi} = b\,.
    \label{eq:dT_dphi}
\end{equation}
This states that nudging a ray by an extra unit of azimuthal deflection $\mathrm{d}\Phi$ increases its travel time by an amount equal to the impact parameter $b$.

The relation in Eq.~\eqref{eq:dT_dphi} is exact for how the time delay is related to the impact parameter and azimuthal sweep. It also provides a convenient route to the coordinate travel time: rather than evaluating the elliptic integral in Eq.~\eqref{eq:elliptic_t}, which involves elliptic integrals of the third kind, we integrate $\mathrm{d}T/\mathrm{d}b = b\,\mathrm{d}\Phi/\mathrm{d}b$ along $b$. No weak-field, small-deflection or thin-lens approximation enters at any stage. 

\subsection{Winding number and relativistic images}
\label{sec:td_winding}
 
So far we have considered the images attached to a single arrival-time surface. In the strong field, however, geodesics can wind around the lens multiple times, producing additional ``relativistic'' images beyond the ``direct'' images of weak-field lensing~\cite{1959RSPSA.249..180D, Ohanian:1987xga, Virbhadra:1999nm, Virbhadra:2008ws, Bozza:2002zj, Bozza:2010xqn}. 
We therefore label each arrival-time surface by a winding number $n$, where $2n$ corresponds to the number of times the geodesic crosses the optical axis before reaching the screen.

An image forms only when the ray sweeps the azimuth required to connect the source to the observer, which for an exactly aligned source and a ray winding $n$ times about the black hole is $\pi + 2\pi n$. Adding this connecting condition to
Eq.~\eqref{eq:generating} as a linear term, we define the generating function per winding number as
\begin{equation}
{\mathcal{I}}_n = \mathcal{I} + (\pi+2\pi n)\,b
= T+ b\left[(\pi+2\pi n)-\Phi\right] \,.
\label{eq:generating_n}
\end{equation}
Using Eq.~\eqref{eq:dT_dphi}, the first two derivatives are simply
\begin{equation}
{\mathcal{I}}_n'(b) = (\pi+2\pi n)-\Phi(b)\,,
\qquad
{\mathcal{I}}_n''(b) = -\Phi'(b)\,,
\end{equation}
such that ${\mathcal{I}}_n'(b)=0$ recovers the connecting condition
$\Phi(b) = \pi + 2\pi n$, and ${\mathcal{I}}_n''$ is fixed entirely by the
curvature of the deflection curve.

Let us now introduce a small displacement of the source  $\beta_{\rm S}\ll 1$. A ray connecting such a source to the observer screen point labeled by $(b,\varphi)$ will, to leading order, accumulate an azimuth of $\pi + 2\pi n - \beta_{\rm S} \cos \varphi$. Combining this with Eq.~\eqref{eq:generating_n}, we define 
\begin{equation}
\mathfrak{t}_{\mathrm{d, n}}(b,\varphi, \beta_{\rm S})
\equiv
\mathcal{I}_n(b)-b\,\beta_{\rm S}\cos\varphi - \mathfrak{t}_{\mathrm{d}}^{\mathrm{ref}},
\label{eq:tsurface}
\end{equation}
At a stationary point, $\varphi=0$ or $\pi$ and $\mathcal I_n'(b)=\beta_{\rm S}\cos\varphi$, so this surface equals the ray travel time $T(b)$, i.e. it recovers the GO images. This is precisely the condition we wanted our time delay surface to satisfy in the strong field. Elsewhere, it is the phase used in the diffraction integral. It accounts for both the difference in accumulated azimuth and coordinate time between different geodesics. 
We choose $\mathfrak{t}_{\mathrm{d}}^{\mathrm{ref}}$ to be the arrival time of the image that arrives first (i.e. the minimum of $\mathfrak{t}_{\mathrm{d, 0}}(b,\varphi)$). This ensures that at the minimum image, $\mathfrak{t}_{\mathrm{d, 0}}(b^\mathrm{ref},\varphi^\mathrm{ref}) = 0$. 

Operationally, to construct $\mathfrak{t}_{\mathrm{d},n} (b,\varphi, \beta_{\rm S})$, we start by obtaining $\mathcal{I}_n (b)$ for an aligned source--lens--observer configuration. This requires solving the integrals in Eq.~\eqref{eq:elliptic_phi} from the observer to the source radius and using Eq.~\eqref{eq:dT_dphi} to find the respective accumulated times. Finally, we include the correction due to an off-axis source $\beta_{\rm S} \cos \varphi$.

The result is that, for each winding number, the strong field time delay surface defined in Eq.~\eqref{eq:tsurface} is a unique scalar function whose stationary structure encodes the geometric-optics images, while also giving direct access to the phase in their neighborhood. This is exactly the stationary phase approximation, which states that in the high-frequency limit, contributions to integrals are dominated by points where the phase is stationary~\cite{Schneider:1992bmb}---the images. Most importantly, the time delay surface gives us access to the local phase near each image. We can also read off its Morse type from the local structure of the generating function: a minimum yields a type-I image with no phase shift, and a saddle a type-II image with a $\pi/2$ shift. In the left panel of Fig.~\ref{fig:td_schw}, we see that we recover the expected stationarity structure of the weak-field case, in that the images form at the minimum and the saddle point of $\mathfrak{t}_{\mathrm{d},0}$. Unless otherwise specified, the results shown in this work correspond to the $n=0$ arrival time surface, as the images that wind additional times around the black hole are exponentially de-magnified~\cite{Bozza:2010xqn}.
%
\subsection{Domain of validity}
\label{sec:td_domain}

We have not yet commented on the size of the observer plane that we consider. The admissible impact parameters form the connecting annulus $b\in(b_c,b_\mathrm{\max}]$, which defines the mask $\mathfrak{D}$: the region of the observer screen whose rays reach the source radius after turning around the lens, and therefore the domain over which the strong-field arrival-time surface is defined and over which we integrate it. The inner edge is the critical impact parameter
\begin{equation}
b_c = 3\sqrt{3}\,M,
\end{equation}
below which rays plunge through the photon ring into the horizon. 
The outer edge follows directly from the turning-point condition $\mathcal R(r_t;b)=0$ in Eq.~\eqref{eq:Rradial}, which gives
\begin{equation}
b^2=\frac{r_t^3}{r_t-2M}.
\end{equation}
For scattering geodesics ($r_t>3M$), the impact parameter increases monotonically with the turning-point radius. Therefore, requiring the ray to reach the source radius, $r_t\leq r_{\rm S}$, gives
\begin{equation}
b\leq b_{\max}(r_{\rm S})
=\sqrt{\frac{r_{\rm S}^3}{r_{\rm S}-2M}}\,,
\end{equation}
with equality corresponding to a ray tangent to the sphere $r=r_{\rm S}$. This expression assumes $r_{\rm O}\geq r_{\rm S}>3M$; reaching the point source also requires the angular connecting condition.

Practically speaking, this means that the construction cannot extend arbitrarily far from, or arbitrarily close to, the optical axis (thus creating the annulus shape seen in the left panel of Fig.~\ref{fig:td_schw}). Additionally, in its current form it cannot capture the effects of retro-lensing, where the source can be lensed even when the source is in between the lens and the observer~\cite{Yu:2021dqx}. Extending the construction to this configuration is left to future work.

\section{The Strong-Field Amplification Function}
\label{sec:amp_f}
In this section, we assemble the arrival-time surfaces of Sec.~\ref{sec:td_surface} into a strong-field amplification function, fix its normalization by matching onto the high-frequency limit, and delimit the range of frequencies over which it can be trusted.
 
Expressed in the ray-tracing framework, the diffraction integral for a point source in the equatorial plane is evaluated per winding number over the region $\mathfrak{D}$,
\begin{equation}
    F_n(\omega,\beta_{\rm S}) = \kappa_\mathrm{n}\omega\int_{\mathfrak{D}} \mathrm{d}^2\vec\theta\; e^{i\omega\,\mathfrak{t}_{\mathrm{d},n}(\vec{\theta})}\,,
    \label{eq:diffrac_rt}
\end{equation}
where $\mathfrak{t}_{\mathrm{d},n}$ is the relative arrival time of Eq.~\eqref{eq:tsurface}, measured relative to its stationary value, and $\kappa_\mathrm{n}$ is a normalization constant fixed below. The total amplification factor is their sum,
\begin{equation}
    F_{\rm RT}(\omega) = \sum_{n} F_n(\omega)\,.
    \label{eq:Ftot}
\end{equation}
Because the higher-winding images are exponentially demagnified~\cite{Bozza:2001xd, Bozza:2002zj}, we present the direct ($n=0$) amplification factor unless stated otherwise.

The normalization of Eq.~\eqref{eq:diffrac_rt} is fixed by demanding agreement with the stationary-phase approximation (SPA) of the diffraction integral, valid in the high-frequency limit where only the stationary points (i.e. the image locations $\vec\theta_j$) contribute,
\begin{equation}
    F(\omega)\approx\sum_j \sqrt{|\mu_j|}\;
    \exp\!\left(i\omega\,\mathfrak{t}_\mathrm{d}(\vec\theta_j) - i\,m_j\,\pi\right),
\end{equation}
where each image carries its own arrival time $\mathfrak{t}_\mathrm{d}(\vec\theta_j)$, magnification $\mu_j$, and Morse index $m_j$ (with $m_j=0$ for a minimum and $m_j=1/2$ for a saddle) respectively. The magnification is defined as the inverse of the determinant of the lensing Jacobian. Following Refs.~\cite{Frittelli:1999yf,Beloborodov:2002mr,Poutanen:2019tcd,brian:prep}, the magnification arbitrarily close to the Schwarzschild black hole can be written as\footnote{This can be seen by inspecting Eq.~(14) of \cite{Poutanen:2019tcd} where we make the identification $\mathcal{D} \rightarrow \mu$, $\psi \rightarrow\pi \pm \beta_S$, and $u \rightarrow 2 M /r$.}
\begin{equation}
    \mu_n(r_{\rm S};b_n) = \frac{b}{r_{\rm S}^2\sqrt{\mathcal R(r_{\rm S};b_n)} \left|\sin \beta_{\rm S}\right|\left|\partial_b \Phi\right|_{b_n}}\,,
    \label{eq:magnification-close-source}
\end{equation}
where this is evaluated at $b_n$, which solves the lens equation for image $n$.

Matching Eq.~\eqref{eq:diffrac_rt} onto this high-frequency limit determines $\kappa_\mathrm{n}$ in terms of the minimum magnification of the images generated per winding number. This ensures that as $\omega \to \infty$, $F_\mathrm{RT} \to F_{\rm SPA}$. The normalization in the spherically symmetric case can be derived analytically (see Appendix~\ref{app:normalization} for details), and takes the form 
\begin{equation}
  \kappa_\mathrm{n} \;=\; \frac{1}{2\pi i}\sqrt{\frac{\mu^\mathrm{ref}\,|\beta_{\rm S}|\,|\mathcal{I}''(b^\mathrm{ref})|}{b^\mathrm{ref}}}\;,
\end{equation}
where $\mu^\mathrm{ref}$ is the magnification of the image at the minimum of the time delay surface (or in other words, the first arriving image), and $b^\mathrm{ref}$ is its impact parameter. We also numerically verify that with this normalization, the magnification of the image at the saddle point of $\mathfrak t_{\rm d}$ is correct, thus validating the approach. This is due to the determinant of the Hessian of the arrival time surface determining the relative amplitudes of the images.

To recover the lower-frequency structure that the SPA misses, where the finite annulus $(b_\mathrm{c},b_{\max}]$ contributes coherently rather than only at its stationary points, we integrate the arrival-time surface directly for a given source--lens--observer configuration.
 
\subsection{Low-frequency limitations}
\label{sec:low_freq}
 
The finite extent of $\mathfrak{D}$ limits the lowest frequency at which the diffraction integral is reliable. The reason is that for a given finite $\mathfrak{D}$, there is a maximum time delay between any two points, say $\Delta \mathfrak{t}_{\rm{d} ,n}^{\rm{max}}$. If the frequency is such that $\omega \, \Delta \mathfrak{t}_{\rm{d},n}^{\rm{max}} \lesssim 1$, then it is easy to see that the time delay surface and the diffraction integral will simply cease to be informative. 
In other words, a trustworthy reconstruction requires the Fresnel zone of a given frequency to lie inside $\mathfrak{D}$. In particular, since sources closer to the lens have reduced $\mathfrak{D}$ (see Sec.~\ref{sec:td_domain}), they have a higher minimum frequency.
Terminating the integral at $b_{\max}$ imposes a hard edge on the screen, which rings in $F(\omega)$ much as a sharply truncated Fourier transform does. The details of our treatment of this edge are given in Appendix~\ref{app:truncation}.

\begin{figure}[t]
    \centering
    \includegraphics[width=\linewidth]{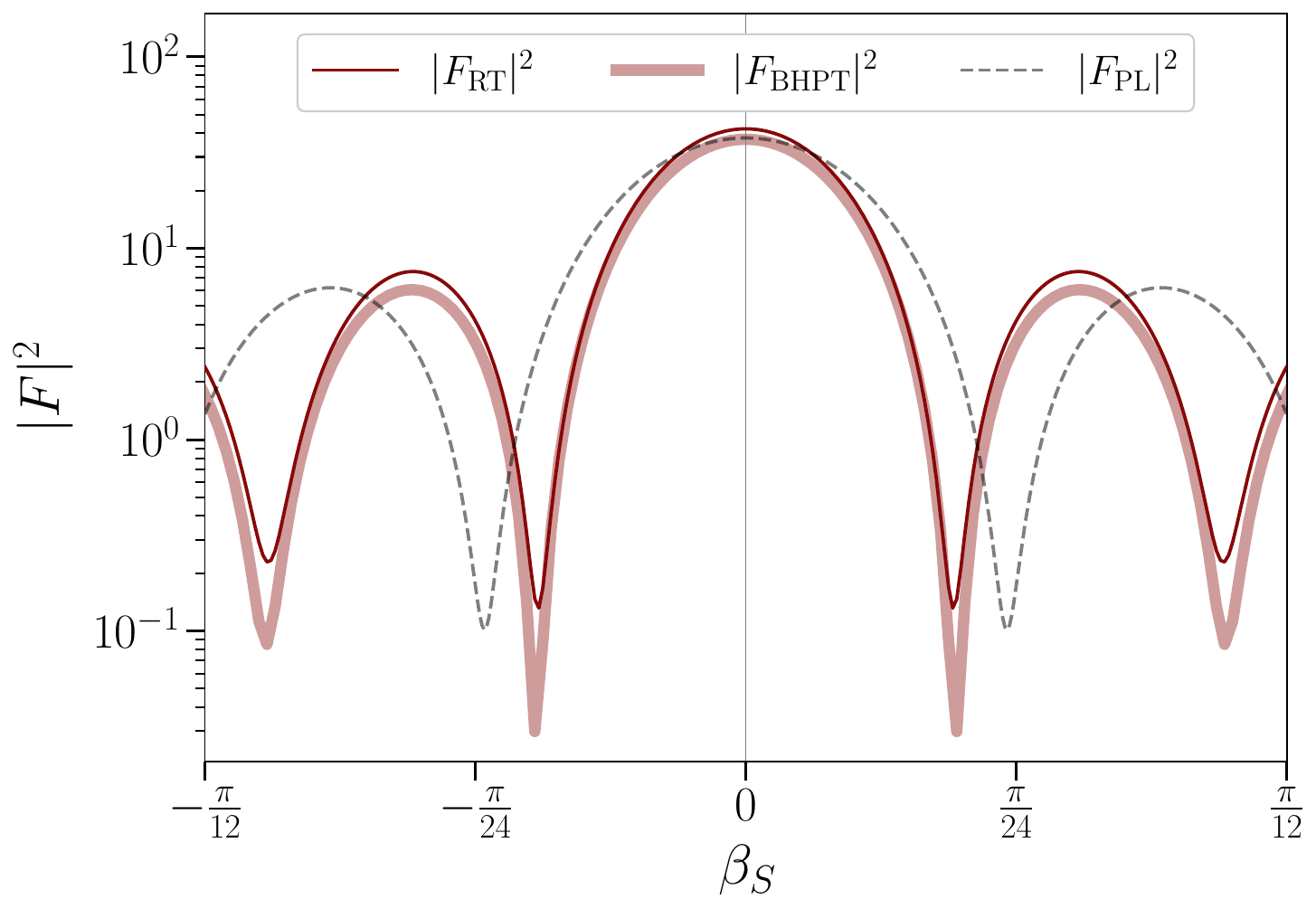}
    \caption{Amplitude of the amplification factor seen by an observer on the equatorial plane, for a source at $r_{\rm S} = 10M$, where $M$ is the mass of the massive Schwarzschild BH.  
    Solid curves indicate results obtained from the black hole perturbation theory model (see Appendix~\ref{app:scalar_WO}) at $M\omega = 3$. The ray tracing  model recovers peak amplification and fringe aperture to 6.5\% and 0.3\% accuracy, respectively. Contrarily, the point lens model is unable to capture the strong field diffraction pattern. 
    }
    \label{fig:F_schwarzschild_SBI}
\end{figure}

Our construction thus does not reproduce the true long-wavelength limit, but it retains sufficient low-frequency resolution to validate against fully relativistic scattering calculations; that comparison is shown in Fig.~\ref{fig:F_schwarzschild_SBI} and explored further in a companion paper~\cite{SieirodosSantos:2026ncf}. 
 
In the opposite regime, Fig.~\ref{fig:td_schw_weak} shows the strong-field amplification function against the point mass lens at large source and observer radii, where the two constructions are expected to agree.

\section{Comparison in the Strong and Weak Field Regimes}
\label{sec:results}

In this section, we compare our formalism against the results of the literature in the strong and weak field regimes. Our points of comparison in the strong field come from the fully relativistic scattering solutions of Refs.~\cite{Santos:2025ass,SieirodosSantos:2026ncf}. By the nature of the construction of their solutions, they are confined to regions where the source is very close to the BH lens ($r_\mathrm{S} \sim 6\text{--}20\, M$), deep in the strong field regime. Refs.~\cite{Santos:2025ass,SieirodosSantos:2026ncf} consider a full quadrupolar GW source, while the amplification function in Eq.~\eqref{eq:diffrac_rt} assumes a scalar spherically symmetric point emitter. For that reason, we compare our results against relativistic scattering from a scalar emitter (see Appendix~\ref{app:scalar_WO}).
However, the rich phenomenology of lensing by a spin-2 source in the strong field is explored in detail in the companion paper~\cite{SieirodosSantos:2026ncf}.

We find remarkable agreement in both the predicted amplitude and the diffraction spacings of the two models. At the peak amplification, we find a $6.5\%$ difference in the values between these two models. Additionally, we find a $0.3\%$ agreement in the locations of the first fringes. This establishes the correspondence between the two models, and clearly illustrates how the formation of images emerges as you take $M \omega \to \infty$. 

It is also worth noting that, even though the comparison is made at a definite frequency, the dependence of this diffraction pattern on this parameter is very straightforward. Indeed, by rescaling the amplification functions by a factor of $\sqrt{M \omega }$ and the angles by $M \omega$, we can get the $F_{\rm BHPT}$ to line up to percent level accuracy, which is explored in detail in the companion paper~\cite{SieirodosSantos:2026ncf}. This is in accordance with the expected scaling in the weak field, and is caused by the universality relations that hold near caustics~\cite{Ezquiaga:2025gkd}. Contrarily, our time delay surface method works better at higher frequencies. Thus, we conclude that, leveraging the universality relations, \emph{our ray tracing model predicts the amplification function in the strong field WO limit for all frequencies $M \omega \gtrsim 1$ to percent level accuracy}.
\begin{figure}[t]
    \centering
    \includegraphics[width=\linewidth]{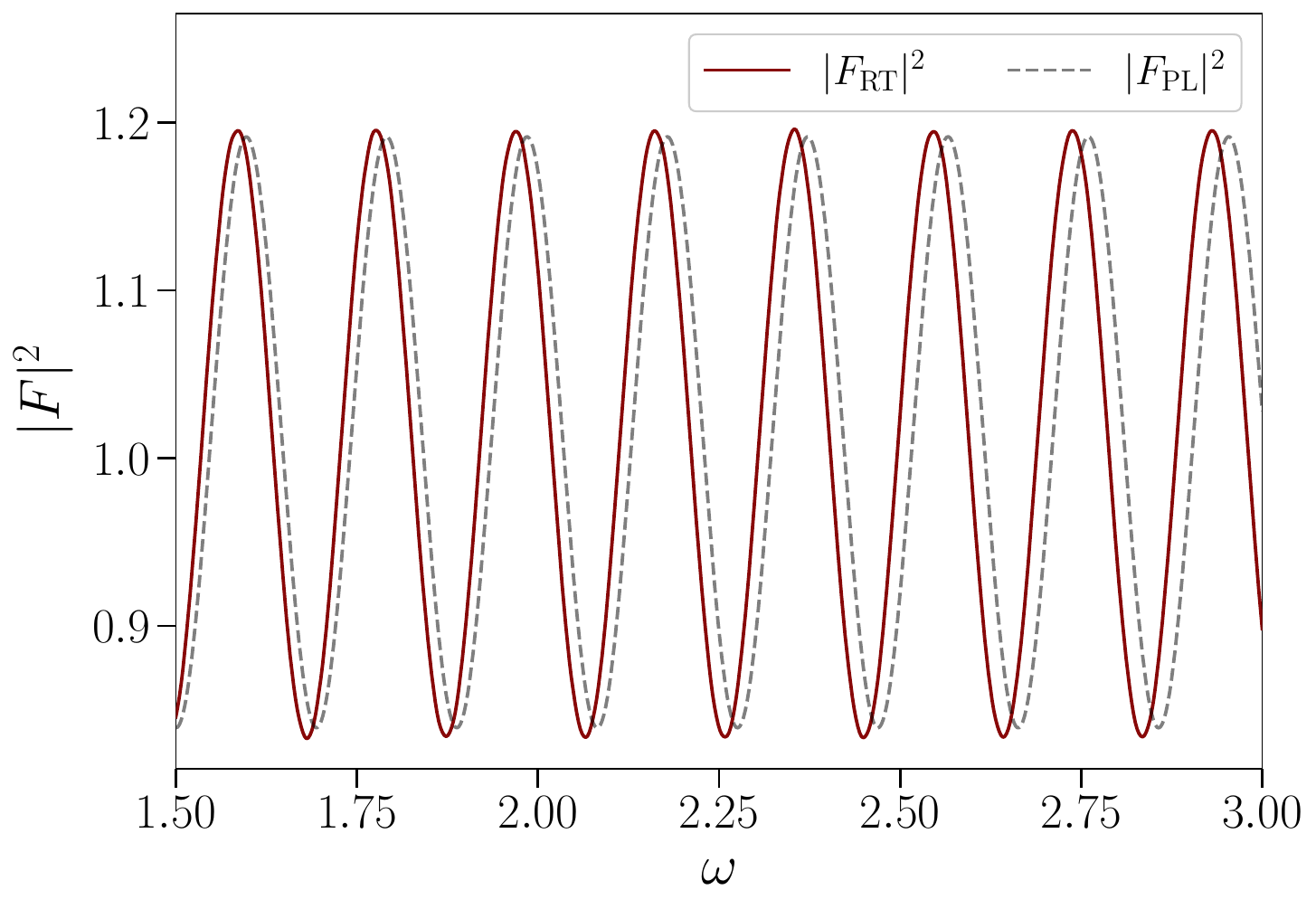}
    \caption{Strong field amplification function compared against the point mass lens in the weak field regime, with $r_{\rm O} =r_{\rm S} = 10^4 M$, $\beta_{\rm S} = \pi/36$. We find the amplitude to match within $<1\%$. The dephasing of the PL comes from its overestimation of the deflection angle, and thus arrival times of images relative the Schwarzschild BH.}
    \label{fig:td_schw_weak}
\end{figure}

 When compared against the PL model, we find a stark difference in both the amplitude and spacing of the lobes of the diffraction pattern, particularly in regions away from the antipode. 
 At the first peak in Fig.~\ref{fig:td_schw_weak}, our model's amplitude agrees remarkably well at $<1\%$. However, there is noticeable dephasing of the peaks of the amplification function, which becomes dramatic at high frequencies.
 This is likely caused by discrepancies between the PL's  estimate of the deflection angles, and thus arrival times, for a source in the strong field regime when compared against a full Schwarzschild lens. 
 Given this disagreement, there are approaches that can be taken to try and estimate an \emph{effective mass} for a PL that can reproduce the diffraction spacing of the strong field results by using the strong field image locations in order to rescale the lens mass (the details of which are explored in Ref.~\cite{SieirodosSantos:2026ncf}). However, we stress that this is only an ad hoc explanation of the difference in the diffraction pattern, and it cannot fully reproduce both the amplitude and fringe spacing predicted by the fully relativistic methods. Additionally, this approach struggles when trying to include additional effects present in this strong field lensing system, such as the effects of the spin-2 nature of the source. 

 As such, we find that our construction is valid from deep within the strong field regime ($r_{\rm S}\sim 6M$), all the way into the weak field regime as $r_\mathrm{S} \to \infty$. More importantly, it provides a smooth connection between these two regimes of lensing. One caveat to the weak field regime of our calculation is that the cost of the computation grows with the distance between the source and the lens ($r_\mathrm{S}$). This is due to the fact that the area of $\mathfrak{D}$ increases as $r_\mathrm{S}$ increases. As such, we need to trace many more rays in order to maintain a consistent phase resolution in our annulus as the source approaches the weak field, increasing the cost of both generating the time delay surface, as well as computing the diffraction integral. Given that for $r_\mathrm{S} < 10^4 M$ there are noticeable differences ($\sim1\%$ difference in amplitude, but with significant dephasing of the peaks at high frequencies) between our model and the PL (already in the most optimistic case where the stellar binary is being modeled as a simple, point source scalar emitter, which we know compact binaries are not), \emph{we recommend the use of the full model for sources within astrophysically favored migration trap distances $r_{\rm S}\lesssim 10^4 M$}.

%

%
\section{Conclusions} \label{sec:conclusions}
%

This work presents a novel method to compute the WO features of strong field lensing near Schwarzschild black holes. The method has been shown to match the results of both fully relativistic wave scattering solutions, as well as the weak field point mass lens model in their respective regimes of validity. Additionally, this construction does not rely on the commonly used thin lens approximation, and can treat source positions arbitrarily close to the innermost stable orbit of the black hole lens. The price to pay for such close alignments however is the lowest valid frequency of the amplification function, which increases as the source approaches the light ring. This construction allows for the first seamless connection between strong field scattering results and the weak field WO, allowing for physically motivated computations of lensing quantities in intermediate regions between the strong- and weak-field regimes. This is a particularly important regime as AGN migration traps theorized to be capable of producing binaries near supermassive black holes are likely to form in these moderately strong gravity regions of $\sim\mathcal{O}(10\text{--}1000)M$~\cite{Samsing:2020tda,Bellovary:2015ifg, Peng:2021vzr, Yunes:2011ws, Pan:2021ksp, Derdzinski:2022ltb}. As such, proper treatment of their lensing signatures could be key in trying to learn about formation channels of binary compact objects~\cite{DOrazio:2017ssb, DOrazio:2019fbq, Leong:2024nnx, Ubach:2025dob, Ubach:2025oxr, Samsing:2024xlo, Samsing:2025rxq, Samsing:2025phj}.

The transparent nature of this formalism allows for clean extensions towards including additional physics in the setup. Notably, the inclusion of the spin-2 nature of GW sources is of great interest, and is explored in a companion paper~\cite{SieirodosSantos:2026ncf}. This entails both taking the emission and intensity profile of the quadrupolar source, as well as taking the polarization content of the emission into account as potential observational signatures of sources being lensed in the strong field.
Another natural extension of this work is to include the spin effects of the black hole lens itself. Astrophysical black holes are almost certainly described by the Kerr family of spacetimes, so this will be a necessary next step in fully modeling these GW sources in realistic environments. Additionally, Kerr introduces interesting phenomenology due to frame dragging effects that allows for exciting and unique signatures of lensing in such environments. 
Finally, the motion of the source is a noteworthy exclusion from this work. While we focused on modeling stationary sources, the model is well suited to include the effects of a binary orbiting a massive BH, which can be traveling at exceptional speeds near the innermost circular orbit of the massive BH. These effects have been found to have considerable effects on lensed waveforms in such systems~\cite{Santos:2025ass}, and are being further explored in an upcoming work~\cite{brian:prep}.

\begin{acknowledgments}
We are grateful to Nicolas Aimar, Conor Dyson, Jose Mar\'ia Ezquiaga, and Vitor Cardoso  for their helpful comments and suggestions.
This work was supported by Research Grants No.~VIL37766 and No.~VIL53101 from Villum Fonden and the DNRF Chair Program Grant No.~DNRF162 by the Danish National Research Foundation.
The Center of Gravity is a Center of Excellence funded by the Danish National Research Foundation under grant No. 184. This project has received funding from the European Union's Horizon 2020 research and innovation programme under the Marie Sklodowska-Curie grant agreement No 101299389 and No 101131233. This work is supported by Simons Foundation International~\cite{sfi} and the Simons Foundation~\cite{sf} through Simons Foundation grant SFI-MPS-BH-00012593-11.
J.S.S. acknowledges financial support provided under the European Union’s H2020 ERC Advanced Grant “Black holes: gravitational engines of discovery” grant agreement no. Gravitas–101052587, and also by the H2020-MSCA-2022-SE
J.S.S also thanks the Danish Data Science Academy for support during the development of this work through Visit Grant 2026-6505, as well as the Fundação para a Ciência e Tecnologia (FCT), Portugal, for the financial support to the Center for Astrophysics and Gravitation (CENTRA/IST/ULisboa) through grant No. UID/PRR/00099/2025 and grant No. UID/00099/2025.
B.C.S. is supported by the European Union's Horizon Europe research and innovation programme under the Marie Sk\l{}odowska-Curie Postdoctoral Fellowship grant agreement No.~101277678 (TripleGW).
Views and opinions expressed are however those of the author(s) only and do not necessarily reflect those of the European Union or the European Research Council. Neither the European Union nor the granting authority can be held responsible for them.
The Tycho supercomputer hosted at the SCIENCE HPC center at the University of Copenhagen was used for supporting this work.
\end{acknowledgments}

\appendix

\section{Derivation of the Normalization Constant}
\label{app:normalization}

We determine the normalization constant $\kappa$ by requiring that, in the high-frequency limit, the diffraction integral reproduces the usual geometric-optics amplitudes of the individual images. For a given winding number $n$, we define the phase
\begin{equation}
    \Psi_n(b,\varphi)
    =
    \mathcal I_n(b)
    -b\beta_{\rm S}\cos\varphi
    -\mathfrak{t}^{\mathrm{ref}},
\end{equation}
where $\mathfrak{t}^{\mathrm{ref}}$ fixes the arbitrary zero of the arrival-time surface. The diffraction integral is then
\begin{equation}
    F_n(\omega)
    =
    \kappa\omega
    \int b\,db\,d\varphi\,
    e^{i\omega\Psi_n(b,\varphi)} ,
    \label{eq:diffrac_norm}
\end{equation}
which can be solved analytically to obtain
\begin{equation}
    F_n(\omega)
    =
    2\pi\kappa\omega
    \int_{b_c}^{b_{\max}}
    b\,J_0(\omega b\beta_{\rm S})
    e^{i\omega[\mathcal I_n(b)-\mathfrak{t}^{\mathrm{ref}}]}\,db .
    \label{eq:F_bessel}
\end{equation}

Equation~\eqref{eq:F_bessel} is convenient numerically. To determine the normalization, however, it is more transparent to return to the two-dimensional form in Eq.~\eqref{eq:diffrac_norm} and take its high-frequency limit using stationary phase.

For $\beta_{\rm S}>0$, the image positions are stationary points of the phase. The azimuthal condition gives
\begin{equation}
    \frac{\partial\Psi_n}{\partial\varphi}=0
    \quad\Longrightarrow\quad
    \varphi_j=0,\pi ,
\end{equation}
while the radial condition gives
\begin{equation}
    \frac{\partial\Psi_n}{\partial b}=0
    \quad\Longrightarrow\quad
    \mathcal I_n'(b_j)
    =
    \beta_{\rm S}\cos\varphi_j .
\end{equation}
The two directions $\varphi_j=0$ and $\pi$ therefore correspond to the two image branches and satisfy different radial equations. The large-argument behavior of the Bessel function in
Eq.~\eqref{eq:F_bessel} contains both of these contributions, so $J_0(\omega b\beta_{\rm S})\simeq1$ cannot be used when taking the high-frequency limit at fixed, nonzero $\beta_{\rm S}$.

Expanding the phase to quadratic order around an image
$(b_j,\varphi_j)$ gives
\begin{equation}
    \Psi_n(b,\varphi)
    \simeq
    \Psi_j
    +
    \frac{1}{2}\mathcal I_n''(b_j)(b-b_j)^2
    +
    \frac{1}{2}b_j\beta_{\rm S}\cos\varphi_j
    (\varphi-\varphi_j)^2 .
    \label{eq:phase_quad}
\end{equation}
The mixed term vanishes because $\sin\varphi_j=0$.

Using the Gaussian integral
\begin{equation}
    \int_{-\infty}^{\infty}
    e^{i\omega a x^2/2}\,dx
    =
    \sqrt{\frac{2\pi}{\omega|a|}}\,
    e^{i\pi\,{\rm sgn}(a)/4}
\end{equation}
in the radial and azimuthal directions, and replacing the screen measure $b\,db\,d\varphi$ by $b_j\,db\,d\varphi$ at leading order, the contribution from image $j$ becomes 
\begin{equation}
    F_n^{(j)}
    \simeq
    2\pi\kappa
    \sqrt{
        \frac{b_j}
        {\beta_{\rm S}|\mathcal I_n''(b_j)|}
    }
    e^{i\omega\Psi_j} e^{i \sigma_j},
    \label{eq:F_spa_intermediate}
\end{equation}

The value of $\sigma_j$ is $i$ for a minimum, and $1$ for a saddle point, and they arise from the local shape of the arrival-time surface around each image. For the Schwarzschild branches considered here, $\mathcal I_n''=-\Phi'(b)>0$. The image at $\varphi_j=0$ curves upwards in both screen directions and is
therefore a minimum, while the image at $\varphi_j=\pi$ curves in opposite directions and is a saddle.

In GO, the corresponding image contributions are
\begin{equation}
    F_{\rm GO}^{(j)}
    =
    \sqrt{|\mu_j|}
    e^{i\omega\Psi_j-i\pi m_j},
\end{equation}
where $m_j=0$ for a minimum and $m_j=1/2$ for a saddle. We fix the normalization using the first-arriving minimum image. Matching its stationary-phase contribution in Eq.~\eqref{eq:F_spa_intermediate} to the geometric-optics amplitude gives
\begin{equation}
    \kappa
    =
    \frac{1}{2\pi i}
    \sqrt{
        \frac{
        |\mu^{\rm ref}|\beta_{\rm S}
        |\mathcal I_n''(b^{\rm ref})|
        }{
        b^{\rm ref}
        }
    } .
    \label{eq:kappa_norm}
\end{equation}

Finally, this derivation assumes isolated images with $\beta_{\rm S}>0$. At exact alignment the images merge into an Einstein ring, so the isolated-image stationary-phase approximation used above no longer applies.

\section{Mitigation of Annulus Truncation Effects}
\label{app:truncation}

For a source behind the lens, the
ray-traced arrival-time surface only exists over a finite range
of impact parameters,
\begin{equation}
    b\in (b_c,b_{\max}] .
\end{equation}
The outer boundary $b_{\max}$ is set by the largest ray that can still
reach the source radius. As a result, the diffraction integral ends
abruptly at this point. This sharp cutoff can introduce artificial
oscillations in the amplification function, in the same way that a
hard edge in an ordinary diffraction problem produces ringing.

After carrying out the azimuthal integral, we write
\begin{equation}
    F_n(\omega)
    =
    \int_{b_c}^{b_{\max}}
    g(b)\,e^{i\omega S(b)}\,db ,
\end{equation}
with
\begin{equation}
    S(b)=\mathcal I_n(b)-t_{\rm ref},
    \qquad
    g(b)=2\pi\kappa\omega\,
    b\,J_0(\omega b\beta_{\rm S}),
\end{equation}
where $J_0$ is a Bessel function of the first kind.

If there is no image close to the outer edge, so that
$S'(b_{\max})\neq0$, the leading contribution produced by the sharp
boundary is 
\begin{equation}
    I_{\rm edge}(\omega)
    =
    \left.
    \frac{
    g(b)\,e^{i\omega S(b)}
    }{
    i\omega S'(b)
    }
    \right|_{b=b_{\max}} .
    \label{eq:edge_term}
\end{equation}
We subtract this term from the numerically evaluated amplification
function in order to suppress oscillations associated with the hard
cutoff at $b_{\max}$. 
This subtraction is equivalent to continuing the
integral beyond the outer edge. Integrating by parts, the contribution
of any smooth extension of the integrand to $b>b_{\max}$, which implicitly assumes that the outer boundary lies away from a stationary point. We therefore check that the recovered
amplification function is stable against changes in the treatment of
the outer edge and in the numerical resolution.

We also note that this prescription only works conveniently in spherically symmetric spacetimes (such as Schwarzschild or Reissner--Nordstr\"om). While a morally similar approach could be taken in Kerr, similar oscillatory artifacts in the final amplification function could be mitigated with appropriate windowing of the edges of the arrival time surface. However, there does not appear to be an obvious choice of the amount of the arrival time surface to taper.

\section{Relativistic scalar scattering model} \label{app:scalar_WO}

Here we outline the most important aspects of the scalar scattering model to produce the strong field comparison in Fig.~\ref{fig:F_schwarzschild_SBI}. The model is identical to what is presented in Refs.~\cite{Santos:2025ass, SieirodosSantos:2026ncf}, except it considers an isotropic source of scalar radiation, rather than a binary sourcing gravitational radiation. 

A massless test scalar field in the Schwarzschild geometry is described by the Klein--Gordon equation
\begin{eqnarray}
    \square_{\rm Sch} \chi(t,r,\theta,\phi) = \mathcal{S}(t,r,\theta,\phi) \, , 
    \label{eq:KG}
\end{eqnarray}
where $\square_{\rm Sch} \equiv \nabla_\mu \nabla^\mu$ is the d'Alembertian of the Schwarzschild geometry, $\chi$ is the scalar field, $(t,r,\theta,\phi)$ are Schwarzschild coordinates, and $\mathcal{S}$ is a source term. For an isotropically emitting point source this is  
\begin{equation}
 \mathcal{S} = e^{i \omega t} \frac{\delta(r-r_{\rm S})}{r^2} \delta(\theta-\pi/2) \delta(\phi-\phi_{\rm S}) \, .
\end{equation}
Plugging this into the equation above, we can easily obtain the resulting scalar field by expanding it and the source into spherical harmonics $Y_{\ell m}(\theta , \phi)$, for $\ell \geq 0$, $-\ell \leq m \leq \ell$. As explained in Ref.~\cite{Santos:2025ass}, the main challenge with this approach is the convergence of the sum over these harmonics, which limits how high $\omega r_{\rm S}$ can be. Choosing, as in the main text, to have the observer at $\phi=0$, the scalar field detected asymptotically is 
\begin{equation}
    \xi(u, r_{\rm O}) \sim \sum_{\ell,m} \frac{A_{\ell m}(\omega) Y^\ast _{\ell m} (\pi/2, \phi_S)}{r_{\rm O}}e^{i \omega u} \, , 
\end{equation}
where $u=t-r_{\rm O}$ is the retarded time and the $A_{\ell m}$ are determined by solving the boundary value problem of Eq.~\eqref{eq:KG}, usually using Green's function methods. In the end the amplification function is simply 
\begin{equation}
    F(\omega, \phi_{\rm S}) \sim \xi / \xi_{\rm Unlensed } = \sum_{\ell,m} A_{\ell m}(\omega) Y^\ast _{\ell m} (\pi/2, \phi_S) \, ,
\end{equation}
where we used $\xi_{\rm Unlensed} \sim e^{i \omega u}/r_{\rm O}$.
%

\end{document}